\documentclass[12pt]{article}
\usepackage[utf8]{inputenc}
\usepackage{amsmath}
\usepackage{amsfonts}
\usepackage{amssymb}
\usepackage{authblk}
\usepackage{xcolor}
\begin{document}
\title{ Spin $\frac{1}{2}$ from Gluons}
\author{A. P. Balachandran}
\affil{Syracuse University, NY 13244-1130}
\date{}
\maketitle

\begin{abstract}
The theta vacuum in $QCD$ is obtained from the standard vacuum, after  twisting by the exponential of the Chern-Simons term. However, a question remains - what is the quantum operator $U(g)$ for winding number $1$?

We construct this operator $U(g)$ in this note. The Poincar\'{e} rotation generators commute with it only if they are augmented by the spin $\frac{1}{2}$ representation of the Lorentz group,  coming from large gauge transformations. This result is analogous to the well-known `spin-isopin'
mixing result due to Jackiw and Rebbi \cite{Jackiw}, and Hasenfratz  and 't Hooft \cite{hasenfratz}. There is  a similar result in fuzzy physics literature \cite{Fuzzy}.  

This shows that  states can drastically affect repreentations of observables. This fact is further shown by charged states dressed by infrared clouds. Following Mund, Rehren and Schroer \cite{MRS}, 
we find that Lorentz invariance is spontaneously broken in these sectors. This result has been extended earlier to $QCD$ (references \cite{nair} given in the Final Remarks) where even the global $QCD$  group is shown to be broken.

It is argued that the escort fields of \cite{MRS} are the Higgs fields for Lorentz and colour breaking. They are string-localised fields where the strings live in a union of de Sitter spaces. Their oscillations and those of the infrared cloud can generate the associated Goldstone modes.    

\end{abstract}

\section{Introduction}

Origin of spin is intimately connected with the Poincare symmetry, as was discovered in Wigner's monumental work on the representation of the Poincar\'e group. One then still feels  an echo of the immortal question of Rabi -"Who ordered that?" when confronted with the existence of fermions. Naively, one can 
just take up the viewpoint that Fermions exist a priori, but the 
early works by Finkelstein and Rubinstein\cite{RubberBand} as well as by Skyrme \cite{Skyrme} showed that the spin-1/2 nature of the fermions can arise because of the non-trivial topology associated with bosonic fields. 

In quantum physics, there are always two related aspects. The first is the algebra $\mathcal{A}$ of
observables which represent elements subject to experimental measurements. The second is 
the state $\omega$ which represents the quantum ensemble which will be subject to measurements.
$\omega$ is a positive linear functional on $\mathcal{A}$, so that if $a \in \mathcal{A}$ , $\omega(a)$ is a 
complex number. Also $\omega (a^* a) \geq 0$ and $\omega(\mathbb{I})=1$,
the two properties needed for a probability measure. 

In this view, the Hilbert space $H$ and the representation of $\mathcal{A}$ on $H$ are emergent concepts  which can be found using the GNS construction. Though the abstract algebra $\mathcal{A}$ remains  the same,
the representations of $\mathcal{A}$ depend on $\omega$.

 It can happen that two $\omega$’s give equivalent
representations, but matrix elements of observables between vectors belonging to these representations
vanish: this vanishing theorem may require taking the direct sum of these representations.

It can also happen that the emergent representations are inequivalent. Here too,
no observable can excite a vector in one to a vector in the other. 

In either case,  we say that the representations are superselected. If a Lagrangian symmetry changes
the superselection sector, it is said to be spontaneously broken. 

In an infinite ferromagnet, the vector states in an irreducible representation can be all 
those with the same direction of  asymptotic spins. Observables can be those which affect the local 
spins without changing the asymptotic value. That defines an irreducible representation of observables.

Another Hilbert space will have vectors with asymptotic direction of spins being in a
different direction, but still observables causing only local disturbances of spins.
These two irreducible representations are equivalent, but no observable has a non-zero
matrix element between vectors of the two representations. 

In the case of a charged 
Higgs field $\phi$ breaking say  gauged $U(1)$ symmetry, it can happen that we have two families
of states defining their Hilbert spaces, the expectation values of $\phi(x)$ as $| \vec{x} |$
goes to infinity differing in magnitude. This difference can be caused by the Higgs potential.
In this case, the $U(1)$ gauge field has different masses in the two cases so that
the representations are inequivalent. But still the local observables define the same
algebra .

These remarks illustrate that we need both the abstract algebra $\mathcal{A}$ and a state on it to
realise $\mathcal{A}$ as operators on a Hilbert space.In this note, we elaborate on this idea for
$SU(N)$ theta vacua in non-abelian gauge theories. These vacua are based on the fact 
that the homotopy group $\pi_3 ( SU(N) ) = Z$, for $N \geq 2$. The quantum states are classified by
representations of this group. The spatial manifold 
is compactified to $S^3$ and  the $N \times N $ matrix  $g_n( \vec{x})$ for $\vec{x} \in S^3$ is valued in a fundamental representation of $SU(N)$.  Here   $g_n(\vec{x}) \in SU(N)$ is a winding number $n$  gauge transformation,
and has the image $\exp( i n \theta)$ on the theta states. They define a representation 
of observables by the GNS construction on these states. If $U(g_n)$ is the quantum operator implementing the winding number $n$ gauge transformation on these theta states, and $g$ 
denotes $g_1$, then $U(g)$ acting on a theta state must have eigenvalue $e^{i \theta}$. 
We will find $U(g)$ explicitly. It is a `large’ gauge transformation so that all 
observables must commute with it. 

The $g$ in question is the configuration that occurs for Skyrme solitons.

For clarification, we add that observables are all the operators  commuting with the complete commuting set (CCS) of large gauge transformations whose eigenvalues label the superselection sector. .They include all small gauge transformations (generated by Gauss law) and all local obsrvables. The small gauge transformations vanish on all the quantum vector states and commute also with the local observables.

\section{Remarks on Gauge Transformations}
We will work with an $SU(N)$ gauge theory with the Gell-Mann matrices $\lambda_\alpha$ as its 
Lie algebra generators in its defining $N$-dimensional representation. We also fix an $SU(2)$ 
subgroup with Pauli matrices $\tau_i$ as its generators.

On a spatial slice $\mathbb{R}^3$, the gauge group $\mathcal{G}$ is the group of smooth maps
$$g:\mathbb{R}^3 \rightarrow SU(N)$$
with $g(x)$ having a definite limit as the spatial coordinate goes to infinity,
that is as $| \vec{x} | \rightarrow \infty$.  It has been called the Sky group by Balachandran 
and Vaidya \cite{sky}. It is the analogue of the Spi group for asymptotically flat spaces introduced
by Ashtekar and Hanson \cite{abhay}.

Let $\lambda_\alpha$ be the $SU(N)$ Gell-Mann matrices. Then if $\Xi$ is a Lie algebra
valued test function, 
$$\Xi(\vec{x} ) = \Xi^\alpha (\vec{x} ) \lambda_\alpha, $$
with  $\Xi^\alpha (\vec{x} )$ approaching definite limits as 
$|\vec{x} | \rightarrow \infty$  ,the Lie algebra generators of the sky group are 
$$ Q(\Xi) = \int d^3 x ~tr~(D_i \Xi (\vec{x})~ E_i ( \vec{x}) + \Xi (\vec{x}) ~J_0 (\vec{x}) )$$
where $D_i$ is the covariant derivative, $E_i$ is the ( Lie algebra valued) electric field,
 $J_0$ is the $SU(N)$ charge density from  matter sources and the trace is in the Lie algebra representation.
 
 Note that 
 \begin{equation}
 [ Q( \Xi), Q(\Xi')] = i Q( [ \Xi, \Xi']).
 \label{commute}
 \end{equation}

If the test functions are compactly supported or vanish fast at infinity, $Q(\Xi)$
represents the smeared Gauss law as one can see by partial integration. So all observables
are required to commute with it. In addition, $Q(\Xi)$ is required to vanish on quantum states.
These are called `small gauge transformations’.

If $\Xi^\alpha$ do not all vanish at infinity, considerations based on locality show
that observables still commute with them \cite{andres}. But $Q(\Xi)$ need no longer vanish 
on quantum states.
For example , in $QED$,  if $\Xi$ goes to a constant at infinity and does not vanish on quantum 
states, then it means that we are working in a charged sector.

So an isometry which does not commute with such $Q(\Xi)$ is not an observable as it changes the superselection sector. It is an intertwiner between two representations of the observables. We 
will see that generic elements of the Lorentz or $SU(N)$ groups do precisely that. Hence these symmetries are spontaneously broken.


\section{The Theta Vacua}
The theta vacua are quantum vector states which respond to $U(g)$ with eigenvalue 
$exp(i\theta)$ and which are invariant under small gauge transformations.They 
can be inferred from instanton physics and are given by
\[|\theta \rangle
 = e^{i \theta \int K(A)} |0  \rangle
\]
where $K(A)$ is the $SU(N)$ Chern-Simons term ,
\[
K(A) = \frac{1}{8 \pi^2} tr~ ( A \wedge dA + \frac{2}{3} A\wedge A\wedge A),
\]
and $|0 \rangle $ is the Poincar\'{e}-invariant vacuum. $|0 \rangle $ remains invariant under the action of both small  and large gauge transformations.

  
Under any winding number 1 transformation $g$ of $A$, $A \rightarrow g A g^{-1} + g dg^{-1}$,\\
$ \int K(A)$
acquires the additional term 
\[
\frac{1}{24 \pi^2 } \int tr ~(dg ~g^{-1})^3 = 1
\]
so that the above Chern-Simons twisted vacuum is indeed the theta vacuum vector.

Note that $\int K(A)$ is invariant under small gauge transformations.

Gauge transformations of Sky for the $SU(3)$ of QCD act on quark fields  $(q_1, q_2, q_3)$ ( suppressing
flavour indices). So if $g$ is an element of the Sky group,  and $U(g)$ is the operator implementing
it, $U(g) q_j( \vec{x},t) U(g)^{-1} = q_k ( \vec{x},t) g(\vec{x})_{kj}$. 


We can guess that $U(g)$ is a large gauge transformation. We propose to show that
it is a finite gauge transforrmation generated by $Q(h)$ where
$h(\vec{x}) := (\vec{\tau} \cdot \hat{\vec{x}}) \tilde{h}(r)$ with
\[
\tilde{h}(0)= 0, \qquad \tilde{h} (\infty ) = -\pi.
\]

Here the Pauli matrices $\tau_i$ are Lie algebra
 generators of the SU(2) acting on the first two quarks.

The above test function is well-defined at $r$ equals 0 as $\tilde{h}$ vanishes there. But as $ \tilde{h}$ is
not zero as $r$ becomes $\infty$, it generates a large gauge transformation.

It will be recognised that $h$ is the winding number 1 Skyrmion configuration. ( See for example \cite{topology_book} ).
An important feature of $h$ is that it is invariant only under the simultaneous rotation of $\hat{\vec{x}}$ and
$\vec{\tau}$. This plays a crucial role in describing spin $\mathit{\frac{1}{2}}$ nucleons using the chiral model of pions.

Let us prove the above claim.

Let $\Psi$ be a coloured field in the $N$-dimensional $SU(N)$ representation. A finite
transformation on $\Psi$ is then given by
\begin{eqnarray}
e^{iQ(h)}\Psi(x)e^{-iQ(h)}&=& \sum_n \frac{i^n}{n!}[Q(h), [Q(h), \cdots[Q(h), \Psi]\cdots ]] 
\nonumber \\
&=&  \sum_n \frac{i^n}{n!} ((\vec{\tau} \cdot \hat{\vec{x}})\tilde{h}(r))^n \Psi((x) = exp(i(\tau \cdot \hat{x}) \tilde{h}(r))\Psi(x) \nonumber \\ &\equiv & g(h)\Psi(x)
\label{what}
\end{eqnarray}

Here $g$ is a Skyrmion configuration which is well-defined :
\[
g(h) = \cos h(r) + i (\tau \cdot \hat{\vec{x}}) \sin h(r).
\]

As $h$ has winding number 1, we have shown that $Q(h)$
is a winding number 1 transformation.

\section{More on Superselection Sectors : Many Theta Vacua}

Let us note a striking feature of theta vacua .When one writes $\vec{\tau}.\hat{\vec{x}}$,
there is an identification of directions such as the third direction in $\vec{\tau}$ and $\vec{x}$ spaces,
or an identification of rotation  generators 
(angular momentum ) in the two spaces. We can also
 write $h’$ equals $(\vec{\tau’}.\hat{\vec{x}})\tilde{h}$ where $\tau'_i$s  are any rotated Pauli matrices. 
 
That too will give a theta sector from its $h$.
But $(h’- h)$ does {\em not} vanish at infinity and so $Q(h’-h)$ is a large gauge transformation. But $Q(h)$ and $Q(h')$ commute as one can show using their 
commutator in eq(\ref{commute})
Hence $Q(h’)$ and $Q(h)$ define different superselection operators even though their eigenvalues on the Chern-Simons-twisted vacua are the same ! 

This has a physical consequence: as discussed in the 
next section, the added term to the orbital angular 
momentum $L_i$ to get the total angular momentum 
compatible with the superselection sector changes 
from $Q(-i \tau_i /2 )$ to $Q(-i \tau'_i /2 )$. These 
act on different SU(2) doublets of gluon octet. The same goes for the curvature octet.

  {\em This is like the situation in a ferromagnet when the 
spins located at the points at infinity are in different directions.The algebras of observables in the two cases are
isomorphic : the isometry is provided by the rotation of spins from one direction to the other.}

If the orbital angular momentum of the gluon $L_i$ , we show in section 7 that the implementable angular momentum $J_i$ and $J'_i$ in the two superselection 
sectors are $L_i + Q( -i \tau_i /2 )$ and $L_i + Q( -i \tau'_i /2 )$. Superposition of such theta vacua states with the `same' angular momentum $J_3$ and $J'_3$  produces a mixed state for the observables. If $CP$  violation from instantons is found, we can ask which mixed state is responsible for it.


\section{ Spin 1/2 from Gluons}
There is a paper by Friedman and Sorkin \cite{sorkin}  with a similar title and 
we have adapted our title from theirs. There are
also papers with similar results  by Jackiw and Rebbi \cite{Jackiw} and  Hasenfratz and ’t Hooft \cite{hasenfratz} in the theory of non-abelian monopoles.

As emphasised in the introduction, quantum theory requires both an algebra of observables and a 
state .( That is the case also in classical theory.) In functional integral approaches, the latter
is defined by the Lagrangian. It can happen that the latter is defined entirely
by bosonic variables, but still quantum theory contains spinorial states. There are 
plenty of examples. The books \cite{topology_book} and \cite{bundle_book} describe many such instances,
both from  soliton physics ( e.g. Skyrmions ) and from  molecular physics ( such as the ethylene
molecule). The theta states are other examples. A  vector state in this case is defined by
the vacuum twisted by a Chern-Simons term. The algebra of observables is gauge invariant.

A superselection sector contains a large gauge transformation $U(g)$.  We claimed above that
this $U(g)$ for us  generates a winding number $1$ transformation. We also claimed that this $U(g)$  
is given by the Skyrmion configuration  for $g$. Let us prove this result.\

\section*{\it A Remark}
Let $h’$ be defined using $\tau’$. The $g(h)$ above and a $g(h’)$ are both 
$-\mathbb{I}$ at $r \rightarrow \infty$ although
$Q(h-h’)$ does not come from the Gauss law and need not vanish on
 quantum states. Thus when restricted to the sphere at $\infty$, the map
from the Lie algebra to the Lie group level is not injective. This result has 
played a role in the above analysis.

Let us return to the main theme. The expression (\ref{what}) is valid also in the {\em pure} 
gluon sector when the state is given by the Chern-Simons-twisted vacuum. The
latter involves the connection $A= A^\alpha \lambda_\alpha$ and $U(g)$ gauge
transforms it with $g(h)$ as in (\ref{what}).

Now the gluons normally rotate only with tensorial angular momentum ( $2\pi$ rotation $=+ \mathbb{I}$ ).  This
operator rotates just $\hat{\vec{x}}$ in $\hat{h}$. But that changes $Q(h)$, 
changing also the superselection sector. We can thus conclude that the canonical angular
momentum $L_i$ for the gluon sector is spontaneously broken.

But consider adding the gauge rotation $Q(\mathbb{I}~ \tilde{h}(r) \tau_i/2)$ to $L_i$. where $\mathbb{I}$ is the constant function with value 1 on $\mathbb{R}^3$  and let us choose the vector state
 $|\theta> \otimes (a,b,0), |a|^2 + |b|^2 =1$. The added term 
rotates $\tau_i$ as well in $Q(h)$ so that $L_i + Q(\mathbb{I}~\tilde{h}(r)~ \tau_i/2)$ commutes with
$Q(h)$ : it does not change the supeselection sector. That is, the total angular
momentum $J_i = L_i + Q(\mathbb{I}~ \tilde{h}(r) \tau_i/2 )$ does not change the superselection
sector.

The $2\pi$ rotation from $J_i$ acting on the above twisted vacuum state changes its sign : the $SU(2)$
Chern-Simons twisted vacuum is spinorial. In this way we get spinorial states in the
 gluon sector.

{\it  If we had considered the vector state  $|\theta > \otimes (a,b,c)$ where the second factor carries the colour representation of quarks and $|a|^2 + |b|^2 + |c|^2 =1$, 
the first two quarks, transforming by $\tau_i/2$, the spin $1/2$ representation of $SU(2)$, become bosonic,  while the third stays fermionic. This has many phenomenological consequences which can be used to test for theta vacua.}

 We will return to this issue in a later work.
 
\section{The Lorentz Group}
Let $K_i$ be the canonical boost associated to $L_i$. Then $K_i + Q( i \mathbb{I}~ \tau_i/2)$ and
$L_i + Q(\mathbb{I}~ \tau_i/2)$ fulfil the $SL(2, C)$ algebra and are appropriate generators for a 
Majorana field.

( We can also consider $L_i + Q( -i \tau_i/2)$ ). A Majorana field transforming unitarily by
these operators can also be constructed using Weinberg’s methods \cite{weinberg}.) 

Unfortunately this choice 
of boosts does not seem to preserve the superselection sector. For example , in $Q(h)$, $\tau_i$
will transform by the non-unitary $(1/2,0)$ representation of $SL(2,C)$ and that does not seem
to be compensated by the transformation of $x$. So the Lorentz group is spontaneously broken,
a result known from other papers. But the spinorial cover of the Euclidean group with $L_i +  
Q(\mathbb{I}~ \tau_i/2)$ and spacetime translations seem implementable in the theta sectors.

In a subsequent paper \cite{next}, we show that infrared effects canonically induce fields on the two-sphere at `infinity'  with covariant $SL(2,C)$. Acting on the vacuum, they create states on the local  algebra which under $SL(2,C)$ intertwine inequivalent irreducible representations.
 

\section{The Chern-Simons term for $SO(3) \subset SU(N)$}
When $N \geq 3$, there is an $SO(3)$ subgroup in $SU(N)$ acting say on the first three components
of the $N$-dimensional vector space. This group had a prominent role in our work on dibaryons \cite{lizzi}
as solitons.

The image of $\tau_i / 2$ in the three-dimensional $SO(3)$ representation is the $3\times 3$ angular matrices $l_i$. 
( These are conventionally called $\theta_i$ as in our group theory book \cite{group_book} , but we will use
$l_i$ instead to  avoid confusion with the theta of theta vacua.) Accordingly, the
Skyrmion configuration is changed to
\[ 
 h'( \vec{x} ) = ( \vec{\mathbf{l}}\cdot \hat{\vec{x}} ) \tilde{h}(r).
\]
 Its finite transformation equals
\[
\hat{g}(\hat{\vec{x}}) = e^{i\vec{l}\cdot\hat{\vec{x}}} h(r)
\]
with winding number 4 and so the eigenvalue of winding number transformation on the
Chern-Simons twisted vacuum is $e^{ 4i\theta}$. The periodicity in theta now is accordingly
$\frac{2 \pi}{4}$.

The angular momentum $J_i = L_i +l_i$ is now tensorial. The boost generators
are $K_i+ Q( i l_i)$ , but the associated Lorentz group changes the superselection sector.

\section{Brief Remarks on Escort Fields}

	We add this brief para to draw attention to the remarkable developments in the theory of string-localised quantum fields and their escort fields, They have a bearing on the results obtained in this paper too. 
	
	In the abstract, we remarked that the Goldstone modes of Lorentz symmetry breaking are incorporated in the escort fields of \cite{MRS}. That is the case : these fields incorporate a 'string' from the direction of the Wilson line, and it can locally fluctuate creating quantised Goldstone modes. But to keep this paper focused,  we will discuss such points in later work\cite{next}.

\section{Final Remarks}
There is more to be said on superselection sectors and their relation for example to
Wilson lines and the Rindler space. They will be discussed in later work.

An older result discussed in \cite{nair} concerns $QCD$ :As it is non-abelian, its  generators do  not generically commute with $Q(\Xi)$ :
 only the stability group of $Q(\Xi)$ does so.

A particular result among others with direct application is the calculation of the Landau-Yang process, strictly forbidden by Lorentz invariance and allowed by its breaking. This calculation with Asorey, Babar, Balachandran, Momen and Qureshi\cite{z2gamma} is completed and also been reported. 

It is striking that theta vacua can convert the gluon sector to spinorial states and that the theta states are infinitely degenerate. These results will have an impact on axion phenomenology, which is 
yet to be explored.

 \section*{Acknowledgments}
I thank Manolo Asorey, Andres Reyes-Lega, V.P.Nair. Sasha Pinzul  and Sachin Vaidya for many discussions. 
I am also grateful to Arshad Momen for critical comments and help with the preparation of this article.


\begin{thebibliography}{99}

\bibitem{RubberBand} D. Finkelstein and J. Rubinstein, J. Math. Phys. 9, 1762–1779 (1968).

\bibitem{Skyrme} T H R Skyrme, Nuclear Physics
{\textbf 31}(1962)556.

\bibitem{Jackiw} R. Jackiw and C. Rebbi, Phys. Rev. Lett. {\bf 36} (1976) 1116.

\bibitem{hasenfratz} P. Hasenfratz and G. 't Hooft, Phys.Rev.Lett. {\bf 36} (1976) 1119.

\bibitem{Fuzzy} A. P.  Balachandran, S.Kurkcuoglu  and S. Vaidya, {\it  Lectures on Fuzzy and Fuzzy Susy Physics}, World Scientific  Publishing Company (2007). 

\bibitem{sorkin} J. L. Friedman and R. D. Sorkin, Phys. Rev. Lett. 44 (1980) 1100. 

\bibitem{MRS}  J. Mund, K. H. Rehren and  B. Schroer, 
arXiv:hep-th/2109.10342.

\bibitem{nair} A.P. Balachandran, V. P. Nair, A. Pinzul, A. F. Reyes-Lega, S. Vaidya, ``Superselection, Boundary Algebras and Duality in Gauge Theories"
( arXiv:2112.08631 [hep-th] ) and refs. 1, 12 and 13 therein. 

\bibitem{sky} A. P. Balachandran  and S. Vaidya, Eur. Phys. J. Plus {\bf 128}  (2013) 118.

\bibitem{abhay} A. Ashtekar and R.  O. Hanson , J. Math. Phys. 19(1978)1534.

\bibitem{topology_book} A. P. Balachandran, G Marmo, B S Skagerstam and A Stern , {\it  Classical Topology and Quantum States. } World Scientific  Publishing Company (1991).

\bibitem{bundle_book}A. P. Balachandran, G Marmo, B S Skagerstam and A Stern,  ``Gauge Symmetries and Fibre Bundles - Applications to Particle Dynamics”, Lecture Notes in Physics 188 , Springer-Verlag(1983), for an updated version see arXive: quant-ph/1702.08910v2. 

 


\bibitem{andres} A. P. Balachandran and A. F. Reyes-Lega in  G. Marmo, D. Martin de Diego, M. Mu\~{n}oz  Lecanda (eds.) Classical and Quantum Physics, Springer Proceedings in Physics. Vol. 229. Springer, Cham.(2019) (https://arxiv.org/abs/1807.05161v2).   


\bibitem{weinberg} S. Weinberg, The Quantum Theory of Fields, Vol. 1. , Cambridge Univ. Press. (2005).  

\bibitem{lizzi} A.P. Balachandran, A. Barducci, F. Lizzi, V.G.J. Rodgers and  A. Stern, Phys.Rev.Lett. 52 (1984) 887, A.P. Balachandran, F. Lizzi, V.G.J. Rodgers and  A. Stern,  Nucl.Phys.B 256 (1985) 525, See also \cite{topology_book}.

\bibitem{group_book}  A. P. Balachandran, S. G. Jo and  G. Marmo,  Group Theory and Hopf Algebras : Lectures for Physicists,
World Scientific Publishing Co Pte Ltd ( 2010). 

\bibitem{next} A.P. Balachandran and colleagues , in preparation.


\bibitem{z2gamma} M. Asorey, A. P. Balachandran, A. Momen and B. Qureshi, arXive: hep-ph/2304.11008. 

\end{thebibliography}
\end{document}